\documentstyle[aps,amssymb,pre,epsfig]{revtex}

\begin{document}

\author{Dmitry E. Pelinovsky \cite{add:dima}  and  Yuri S. Kivshar}
\address{Optical Sciences Centre, Research School of  Physical Sciences and Engineering \\  The Australian National University, Canberra, ACT 0200, Australia}

\title{Stability criterion for multi-component solitary waves}

\maketitle

\begin{abstract}
We obtain the most general matrix criterion for stability and instability 
of {\em multi-component solitary waves} considering a system 
of $N$ incoherently coupled nonlinear Schr\"{o}dinger equations. 
Soliton stability is studied as a constrained variational 
problem which is reduced to finite-dimensional linear algebra. 
We prove that unstable (all {\em real} and {\em positive}) 
eigenvalues of the linear stability problem for multi-component 
solitary waves are connected with negative eigenvalues of the 
Hessian matrix, the latter is constructed for the energetic 
surface of $N-$component spatially localized stationary solutions.
\end{abstract}


\section{Introduction}

Recent discovery of self-focusing of partially coherent light and 
experimental observation of the so-called {\em incoherent spatial 
solitons} \cite{incoh} has called for a systematic analysis of the 
properties and stability of {\em multi-component} and {\em multi-parameter} 
solitary waves. Incoherent solitons are generated in 
noninstantaneous nonlinear media such as biased photorefractive 
crystals. In this case, a self-consistent modal theory 
\cite{modal}, which is equivalent to the coherent density approach, 
describes the incoherent solitons with the help of a system of coupled nonlinear 
Schr\"odinger (NLS) equations (see also \cite{other,exact,2d}). 
Similar models appear, in a different physical context,
in the theory of soliton wavelength-division multiplexing \cite{wdm},
multi-channel bit-parallel-wavelength optical fiber networks \cite{bpw},
multi-species and spinor Bose-Einstein condensates \cite{bec}, and
other important applications \cite{applic}. In all such physical models 
solitary waves are {\em multi-component}, being described by 
localized solutions of the coupled nonlinear equations. In some 
very special cases, the coupled system allows for explicit analytical 
solutions (see, e.g., Ref.  \cite{exact}) but, generally speaking, the nonlinear 
models with multi-component solitary waves are {\em nonintegrable}. 
Stability of solitary waves is therefore a crucial issue for any 
kind of their applications.

The study of soliton stability has a long history. Stability of 
one-parameter solitary waves has been already well understood 
for both fundamental (single-hump and nodeless) solitons 
\cite{VK,ZK,Weinstein} and solitons with nodes and 
multiple humps \cite{jones,Gr}. The pioneering results 
of Vakhitov and Kolokolov \cite{VK} found their rigorous 
justification in a general mathematical theory of  
Grillakis, Shatah, and Strauss \cite{Gr1}.  Although the 
corresponding stability and instability theorems for the scalar 
NLS models extend formally to the case of multi-parametric 
solitons \cite{Gr1}, all the cases analyzed so far correspond 
to solitary waves with effectively {\em a single parameter}.  

A recent progress in the study of soliton instabilities is associated with
 the application of a bifurcation theory valid for {\em weakly 
unstable}  stationary localized waves. In this method, the corresponding 
unstable eigenvalue of the associated spectral problem is treated 
as a small parameter of multi-scale asymptotic expansions \cite{P1}. In the 
case of multi-parameter solitary waves, a simplified version 
of this method is usually reduced to a number of 
``magic determinants'' constructed from the derivatives of 
the system invariants near a marginal stability line 
\cite{P3,three,multi,skryabin}.  However, such a bifurcation 
method {\em has no rigorous proof}, and it does not allow to 
predict the complete domains of the soliton stability and instability, 
since more general {\em   oscillatory instabilities} may occur as well 
\cite{Gr,gap,coupled2}.

In this paper, we present {\em a complete theory} for stability and 
instability  of {\em multi-parameter solitary waves} considering a 
particular example of a system of $N$ incoherently coupled 
NLS equations. Our results include the asymptotic bifurcation 
method with the determinant criterion as {\em a simple 
near-threshold limiting case}. They also expand the applicability 
boundaries of the previously known mathematical theorems \cite{Gr1} 
to the case of multi-component and multi-parameter solitary waves. 

The system of incoherently coupled NLS equations has already been 
studied in many papers  (see, e.g., Refs. \cite{berge,Yew,Yang} 
to cite a few). 
However, the study of stability  of  single-hump and multi-hump 
solitary waves was restricted again by a single-parameter case, 
when soliton components have a similar shape and their amplitudes 
are proportional to each other \cite{berge}. In this paper, we expand 
those results and present, for the first time to our knowledge, a complete matrix 
analysis of the  constrained variational problem leading to the finite-dimensional 
linear algebra. Although some of our results depend on the properties 
which are specific to the model under consideration, we believe that both 
the method and the basic results can be generalized, under proper assumptions, 
to be applied to other types of  nonlinear physical models that support 
multi-parameter solitary waves. 

\section{Model and Basic Results}

We consider nonlinear interaction of $N$ optical modes that 
describe either a multi-mode structure of a partially 
incoherent self-trapped beam or incoherent coupling between 
optical channels with different wavelengths in a fiber.  Then, 
the amplitude envelopes of partial modes satisfy the following 
system of incoherently coupled NLS equations,
\begin{equation}
\label{NLS}
i \frac{\partial \psi_{n}}{\partial z} + d_n \nabla^2_{\bf x} \psi_{n}
+ \left( \sum_{m=1}^N \gamma_{nm} |\psi_m|^2 \right) \psi_n = 0,
\end{equation}
where $\nabla_{\bf x}^2$ stands for Laplacian in the 
$D-$dimensional space ${\bf x} = (x_1,...,x_D)$,  and 
all the coefficients $d_n$ are assumed to be positive. 
When one of the variables of the vector ${\bf x}$ stands for 
time, Eqs. (\ref{NLS}) describe the spatio-temporal dynamics 
of self-focused and self-modulated light in  the form  of the so-called {\em light bullets}. 

Provided the symmetry conditions $\gamma_{nm} = \gamma_{mn}$ 
are satisfied, the system (\ref{NLS}) conserves the Hamiltonian,
\[
H = \int_{-\infty}^{\infty} d {\bf x} \left( \sum_{n=1}^N d_n
\left| \nabla_{\bf x} \psi_{n} \right|^2 - \frac{1}{2}
\sum_{n=1}^N \sum_{m=1}^N \gamma_{nm} |\psi_n|^2 |\psi_m|^2 \right),
\]
the individual mode powers,  
$Q_n = \frac{1}{2} \int |\psi_n|^2 d {\bf x}$, 
and the total field momentum. Localized solutions of Eqs. (\ref{NLS}) 
for the fundamental solitary waves are defined as 
$\psi_n = \Phi_n({\bf x}) e^{i \beta_n z}$,  where $\Phi_n({\bf x})$ 
are real functions {\em with no nodes},  and $\beta_n$ are 
{\em positive} propagation constants. The 
soliton solutions are stationary points of the Lyapunov functional,
\begin{equation}
\label{Lyp}
\Lambda[\mbox{\boldmath $\psi$}] = H[\mbox{\boldmath $\psi$}] +
\sum_{n=1}^N \beta_n Q_n[\mbox{\boldmath $\psi$}],
\end{equation}
i.e. the first variation of $\Lambda[\mbox{\boldmath $\psi$}]$ 
vanishes at $\mbox{\boldmath $\psi$} = {\bf \Phi}({\bf x})$. 
The second variation of $\Lambda[\mbox{\boldmath $\psi$}]$ 
defines the stability properties: negative directions of 
the second variation corresponds to unstable eigenvalues 
in the soliton stability problem (see, e.g.,  Ref. \cite{ZK} for 
a review of the basic results).

The stability problem is defined by minimizing the second variation
of the Lyapunov functional $\Lambda[\mbox{\boldmath $\psi$}]$,
\begin{equation}
\label{second}
\delta^2 \Lambda = \int_{-\infty}^{\infty} d {\bf x}
\left[ \langle {\bf u} | {\bf L}_1 {\bf u} \rangle +
\langle {\bf w} | {\bf L}_0 {\bf w} \rangle \right],
\end{equation}
where ${\bf u}({\bf x})$ and ${\bf w}({\bf x})$ are perturbations
of the multi-component solitary wave taken in the form
 $\mbox{\boldmath $\psi$} 
= {\bf \Phi}({\bf x})
+ \left[ {\bf u} + i {\bf w} \right]({\bf x}) e^{\lambda z}$, and
the scalar product is defined as $\langle {\bf f} | {\bf g} \rangle =
\sum_{n=1}^N f_n^* g_n$. The matrix Sturm-Liouville operator
${\bf L}_0$ has a diagonal form with the elements
$$
(L_0)_{nn} = - d_n \nabla^2_{\bf x} + \beta_n
- \sum_{m=1}^N \gamma_{nm} \Phi_m^2,
$$
and the matrix operator ${\bf L}_1$ has the elements
$$
(L_1)_{nn} = - d_n \nabla^2_{\bf x} + \beta_n
- \sum_{m=1}^N \gamma_{nm} \Phi_m^2 - 2 \gamma_{nn} \Phi_n^2,
$$
at the diagonal,  and $(L_1)_{nm} = - 2 \gamma_{nm} \Phi_n \Phi_m$, 
off the diagonal. Operators $L_0$ and $L_1$ determine the linear
eigenvalue problem for stability of multi-component solitary waves,
\begin{equation}
\label{eigenvalue}
{\bf L}_1 {\bf u} = - \lambda {\bf w}, \;\;\;\;
{\bf L}_0 {\bf w} = \lambda {\bf u}.
\end{equation}
Both the linear problem (\ref{eigenvalue}) and minimization
problem (\ref{second}) should satisfy a set of $N$ constraints,
\begin{equation}
\label{constr}
F_n = \int_{-\infty}^{\infty} d {\bf x}
\langle \Phi_n {\bf e}_n | {\bf u} \rangle = 0,
\end{equation}
where ${\bf e}_n$ is the $n^{\mbox{th}}$ unit vector, which correspond to the conservation of the individual powers $Q_n$ under the action of a perturbation described by a vector (${\bf u},{\bf w}$).

First of all, we recall the main result of Refs. \cite{VK,ZK,Weinstein} 
that the {\em one-parameter solitary waves with no nodes} ($N=1$) are 
stable in the framework of the constrained variational problem 
(\ref{second})-(\ref{constr}) provided the energetic surface 
$\Lambda_s(\mbox{\boldmath $\beta$}) = \Lambda[{\bf \Phi}]$ is concave up, i.e.
\begin{equation}
\label{crit}
\frac{d^2 \Lambda_s}{d \beta_1^2} = \frac{d Q_1}{d \beta_1} > 0.
\end{equation}
Under this condition, the linear eigenvalue problem (\ref{eigenvalue})  
{\em has no unstable eigenvalues}, i.e. those with  {\em a positive real 
part} $\lambda$. Otherwise, the second variation (\ref{second}) constrained 
by the set (\ref{constr}) has a single negative  direction that corresponds  
to a single positive eigenvalue $\lambda$ in the linear eigenvalue problem 
(\ref{eigenvalue}) \cite{VK,ZK}. 
The stability criterion for scalar (or one-component) NLS solitons 
holds when the self-adjoint operator ${\bf L}_1$ has a single negative 
eigenvalue, i.e. when the second variation (\ref{second}), without 
the constraint (\ref{constr}) imposed, has a single negative direction. 
If the latter condition is not satisfied,  as it happens for solitary 
waves with nodes, the fundamental criterion for the soliton instability 
can be extended only for a special case \cite{jones,Gr}, while more generic 
mechanisms of oscillatory instabilities, associated with complex 
eigenvalues of the linear eigenvalue problem,  may appear beyond the prediction 
of the fundamental criterion \cite{Gr,gap,coupled2}.

Here we extend the soliton stability analysis to the case of multi-component solitary waves described by a system of incoherently coupled NLS equations 
(\ref{NLS}).  We assume that the number of negative directions 
(eigenvalues) of the second variation $\delta^2 \Lambda$ is fixed,  and we
denote it as $n(\Lambda)$. The unstable eigenvalues $\lambda$ of the
linear problem (\ref{eigenvalue}) are connected with some negative
eigenvalues of the matrix ${\bf U}$ defined by the elements
\begin{equation}
\label{Hessian}
U_{nm} = \frac{\partial^2 \Lambda_s}{\partial \beta_n \partial \beta_m}
= \frac{\partial Q_n}{\partial \beta_m} = \frac{\partial Q_m}{\partial
\beta_n}.
\end{equation}
The matrix ${\bf U}$ is the Hessian matrix of the energetic surface
$\Lambda_s(\mbox{\boldmath $\beta$})$. We denote the number of
positive eigenvalues of the matrix ${\bf U}$ as $p(U)$, and 
the number of its negative eigenvalues as $n(U)$, so that 
$p(U) + n(U) \leq N$, since some eigenvalues may be zeros in a degenerate 
(bifurcation) case.  As is shown below, both  $p(U)$ and $n(U)$ 
satisfy some additional constraints,
\begin{equation}
\label{restrict}
p(U) \leq \min\{N,n(\Lambda)\}, \;\;\;\;   
n(U) \geq \max\{0,N-n(\Lambda)\}.
\end{equation}
Within these notations, we formulate (and prove below) the following
fundamental results on stability and instability of multi-component
solitary waves of the coupled NLS equations (\ref{NLS}): \\

(i) the linear problem (\ref{eigenvalue}) may have at most $n(\Lambda)$ 
unstable eigenvalues $\lambda$, all {\em real} and {\em positive}; \\

(ii) a multi-component soliton is {\em linearly unstable} provided 
$p(U) < n(\Lambda)$; then the linear problem (\ref{eigenvalue}) has 
$n(\Lambda) - p(U)$ real (positive or zero-becoming-positive) eigenvalues $\lambda$; \\

(iii) a multi-component soliton is {\em linearly stable} provided
$p(U) = n(\Lambda) (\leq N)$; in the case  $n(\Lambda) = N$ this criterion implies that the energetic surface  $\Lambda_s(\mbox{\boldmath $\beta$})$ is concave up in the $\mbox{\boldmath $\beta$}$-space; \\

(iv) a single eigenvalue $\lambda$ crosses a marginal stability curve
when the matrix ${\bf U}$ possesses a zero-becoming-negative 
eigenvalue; the normal form for the instability-induced dynamics of 
multi-component  solitary waves resembles the equation of motion 
for an effective classical  particle subjected to a $N$-dimensional potential field, 
\begin{equation}
\label{particle}
E = \frac{1}{2} \sum_{n=1}^N \sum_{m=1}^N M_{nm} \frac{d \nu_n}{d z}
\frac{d \nu_m}{d z} + W(\mbox{\boldmath $\beta$},\mbox{\boldmath $\nu$}),
\end{equation}
where $M_{nm}$ are the elements of the positive-definite ``mass matrix''
[see Eq. (\ref{mass}) below], $\mbox{\boldmath $\nu$}$ is the vector describing a perturbation  to the soliton parameters $\mbox{\boldmath $\beta$}$, and $W(\mbox{\boldmath $\beta$},\mbox{\boldmath $\nu$})$
is an effective potential energy defined as
\begin{equation}
\label{potential}
W(\mbox{\boldmath $\beta$},\mbox{\boldmath $\nu$}) =
H_s(\mbox{\boldmath $\beta$} + \mbox{\boldmath $\nu$}) -
H_s(\mbox{\boldmath $\beta$})
+ \sum_{n=1}^N \left( \beta_n + \nu_n \right) \left[ Q_{sn}(\mbox{\boldmath
$\beta$}
+ \mbox{\boldmath $\nu$}) - Q_{sn}(\mbox{\boldmath $\beta$}) \right].
\end{equation}

These results should be compared with the results following from the  stability and instability theorems earlier formulated by Grillakis, Shatah, and Strauss 
\cite{Gr1}. The stability result (iii), i.e. the condition 
$p(U) = n(\Lambda)$,  is {\em identical} to that of  the stability theorem \cite{Gr1}, but the instability results (i)-(ii) 
are {\em more general and explicit}. In particular, the theorem of Grillakis {\em et al.}   \cite{Gr1} guarantees the soliton instability provided the difference  $n(\Lambda) - p(U)$ is odd.  However, our results predict 
that the soliton instability always occurs for 
$n(\Lambda) - p(U) > 0$,  being associated with exactly 
$n(\Lambda) - p(U)$ non-negative real eigenvalues $\lambda$ 
of the linear eigenvalue problem (\ref{eigenvalue}). Moreover, 
according to our result (iv), each new unstable eigenvalue 
$\lambda$ appears via {\em a bifurcation} at the marginal stability curve
 where the determinant of the matrix ${\bf U}$ vanishes, i.e. it is connected 
with a zero-becoming-negative eigenvalue of the Hessian matrix ${\bf U}$.  
If $n(\Lambda) > N$, unstable eigenvalues $\lambda$ 
originated from the negative eigenvalues of the Hessian matrix 
${\bf U}$ co-exist with  $n(\Lambda) - N$  unstable eigenvalues of the linear problem (\ref{eigenvalue}),  i.e. a solitary wave is unconditionally unstable when  $n(\Lambda) > N$. 

\section{A Proof of the Basic Results}

Here we develop the analysis of the problem (\ref{second})-(\ref{constr}), 
in order to prove the results (i)-(iv) formulated above.  The Sturm-Liouville 
operators $(L_0)_{nn}$ are all non-negative since the fundamental (node-less) 
solution ${\bf \Phi}({\bf x})$ for a solitary wave reaches a bottom of the 
spectrum at zero: $(L_0)_{nn} \Phi_n = 0$. As a result,
\begin{equation}
\label{decomp}
\mbox{(min)     } \delta^2 \Lambda = \int_{-\infty}^{\infty} 
d {\bf x} \; \langle {\bf u} | {\bf L}_1 {\bf u} \rangle = 
\sum_{\mu} \mu \int_{-\infty}^{\infty} d {\bf x} \; \langle 
{\bf u}_k | {\bf u}_k \rangle.
\end{equation}
Here ($\mu$,${\bf u}_k$) are eigenvalues and eigenfunctions of
the auxiliary linear problem,
\begin{equation}
\label{auxillary}
{\bf L}_1 {\bf u}_k = \mu {\bf u}_k - \sum_{m=1}^N \nu_m
\Phi_m({\bf x}) {\bf e}_m.
\end{equation}
The linear problem (\ref{auxillary}) is constrained by the set 
(\ref{constr}) and the parameters $\nu_1$, $\nu_2$,...,$\nu_N$ 
have the meaning of the Lagrange multipliers.

Let us suppose that the Sturm-Liouville matrix operator ${\bf L}_1$ has
$n(\Lambda)$ negative eigenvalues $\mu = \{ \mu_{-n(\Lambda)},
\mu_{-n(\Lambda) + 1}, ..., \mu_{-1} \}$ corresponding to the
 eigenfunctions
${\bf u} = \{ \mbox{\boldmath $\psi$}_{-n(\Lambda)}({\bf x}),
\mbox{\boldmath $\psi$}_{-n(\Lambda) + 1}({\bf x}), ...,
\mbox{\boldmath $\psi$}_{-1}({\bf x}) \}$;  a single zero eigenvalue
with a one-node eigenfunction ${\bf u} = d {\bf \Phi}/dx$ and
the rest of the spectrum is positive and contains $N$ branches
of the continuous spectrum,  for $\mu > \{ \beta_1, \beta_2,...,
\beta_N \}$, and some isolated positive eigenvalues,  for
$\mu = \{ \mu_1, \mu_2, ..., \mu_{p} \}$.

The mathematical problem can then be reformulated in the following way.
The linear operator ${\bf L}_1$ has $n(\Lambda)$ negative eigenvalues
that generate negative directions of the second variation $\delta^2 \Lambda$.
However, the corresponding eigenfunctions do not satisfy generally
the constraints (\ref{constr}). By introducing the Lagrangian multipliers 
in (\ref{decomp}) and (\ref{auxillary}), we satisfy a constrained 
minimization problem (\ref{second}) and (\ref{constr}) but, 
due to this procedure,  the number of negative eigenvalues may be reduced.
We will show how to connect the total number of negative
eigenvalues of the constrained problem (\ref{constr}), (\ref{decomp}), and 
(\ref{auxillary}) with the negative eigenvalues of the Hessian matrix (\ref{Hessian}).
But,  as a prerequisite, we prove two additional results for the spectrum of 
the problem (\ref{eigenvalue}): \\

(i) the spectrum of $\lambda^2$ is real, i.e. oscillatory instabilities
are prohibited; \\

(ii) each negative direction ($\mu, {\bf u}_k$) of the problem (\ref{auxillary}) 
generates an unstable (positive) eigenvalue $\lambda$ of the problem (\ref{eigenvalue}). \\

To prove the statement (i), we notice that the matrix operator ${\bf L}_0$ 
can be factorized as ${\bf L}_0 = \sum_{d=1}^D {\bf M}^+_d {\bf M}_d$, 
where ${\bf M}_d$ has a diagonal form with the following matrix elements
$$
\left( M_d \right)_{nn} = \sqrt{d_n} \left[ - \partial_{x_d} + 
\frac{1}{\Phi_n({\bf x})} \partial_{x_d} \Phi_n({\bf x}) \right],
$$
provided the soliton solutions $\Phi_n({\bf x})$ have no nodes
in a finite domain. Using this factorization, the linear
problem (\ref{eigenvalue}) can be rewritten for the function
${\bf u} = \sum_{d=1}^D {\bf M}^+_d {\bf v}_d$ as follows,
$$
\sum_{d' = 1}^D {\bf M}_d {\bf L}_1 {\bf M}_{d'}^+ 
{\bf v}_{d'} = - \lambda^2 {\bf v}_d.
$$
Since the matrix operator with the elements ${\bf M}_d {\bf L}_1 {\bf M}^+_{d'}$ 
is Hermitian,  its eigenvalues ($-\lambda^2$) are all real. \\

To prove the statement (ii), we suppose that we have constructed a 
negative direction ($\mu,{\bf u}_k$)
of the problem (\ref{decomp})-(\ref{auxillary}) subject to the 
constraints (\ref{constr}). Then, the linear problem (\ref{eigenvalue}) 
has an unstable eigenvalue $\lambda$ defined as,
\begin{equation}
\label{connection}
\lambda^2 = - \frac{\langle {\bf u}_k | {\bf L}_0 {\bf L}_1 {\bf u}_k
\rangle}{\langle {\bf u}_k | {\bf u}_k \rangle} = - \mu
\frac{\langle {\bf u}_k | {\bf L}_0 {\bf u}_k \rangle}{\langle
{\bf u}_k | {\bf u}_k \rangle}.
\end{equation}
Since the linear operator ${\bf L}_0$ is  positive definite for any 
${\bf u}_k \neq ({\bf 0}, {\bf \Phi})$, we have  $\lambda^2 > 0$ 
for any $\mu < 0$. 

Our next goal is to construct solutions to the auxillary problem (\ref{auxillary}).
Since the matrix operator ${\bf L}_1$ is Hermitian, it has a complete
spectrum in a Hilbert space that is suitable for expanding the vector
function ${\bf u}_k({\bf x})$. We present such a spectral decomposition
in the form,
\begin{equation}
\label{expansion}
{\bf u}_k({\bf x}) = \sum_{m=1}^N \nu_m \left( \sum_{\mu_r < 0}
\frac{ \langle \mbox{\boldmath $\psi$}_r | \Phi_m {\bf e}_m \rangle}{\mu - \mu_r}
\mbox{\boldmath $\psi$}_r ({\bf x}) + \sum_{\mu_r > 0}
\frac{ \langle \mbox{\boldmath $\psi$}_r | \Phi_m {\bf e}_m \rangle}{\mu - \mu_r}
\mbox{\boldmath $\psi$}_r ({\bf x}) \right),
\end{equation}
where the sum $\sum_{\mu_r < 0}$ contains $n(\Lambda)$ terms from the
negative spectrum, while the sum $\sum_{\mu_r > 0}$ includes
schematically both the discrete and continuous positive spectra of
${\bf L}_1$. The contribution from the neutral eigenfunction
${\bf u} = d {\bf \Phi}/dx$ vanishes due to the symmetry properties.
General solution (\ref{expansion}) has to be constrained by the conditions
(\ref{constr}). This system reduces to the linear algebra for the
Lagrange multipliers, ${\bf A}(\mu) {\bf \nu} = {\bf 0}$, where the
matrix ${\bf A}(\mu)$ has a symmetric form with the elements:
\begin{equation}
\label{algebra}
A_{nm}(\mu) = \sum_{\mu_r < 0} \frac{ \langle \Phi_n {\bf e}_n |
\mbox{\boldmath $\psi$}_r \rangle \; \langle \mbox{\boldmath $\psi$}_r 
| \Phi_m {\bf e}_m \rangle}{\mu - \mu_r} +
\sum_{\mu_r > 0} \frac{ \langle \Phi_n {\bf e}_n |
\mbox{\boldmath $\psi$}_r \rangle \; \langle \mbox{\boldmath $\psi$}_r 
| \Phi_m {\bf e}_m \rangle }{\mu - \mu_r}.
\end{equation}

The linear system ${\bf A}(\mu) \mbox{\boldmath $\nu$} = \gamma
\mbox{\boldmath $\nu$}$ has generally $N$ real eigenvalues
$\gamma_1(\mu)$, $\gamma_2(\mu)$, ..., $\gamma_N(\mu)$.
These eigenvalues are continuous functions of $\mu$ for $\mu \leq 0$, 
except for $n(\Lambda)$ resonant planes at $\mu = \{ \mu_{-n(\Lambda)},
\mu_{-n(\Lambda) + 1}, ..., \mu_{-1} \}$. At these planes, the
matrix ${\bf A}(\mu)$ has poles and the eigenvalues $\gamma(\mu)$
may have singularities. Below, we prove the following three properties
of the eigenvalues $\gamma(\mu)$: \\

(i) all eigenvalues $\gamma(\mu)$ are {\em negative} for
$\mu < \mu_{-n(\Lambda)} (< 0)$; \\

(ii) each eigenvalue $\gamma(\mu)$ is a decreasing function of $\mu$ 
for $\mu \leq 0$,  except for $n(\Lambda)$ resonant planes at
$\mu = \{ \mu_{-n(\Lambda)}, \mu_{-n(\Lambda) + 1}, ..., \mu_{-1} \}$; \\

(iii) at least ($N-1$) eigenvalues $\gamma(\mu)$ are continuous 
at any of the resonant planes $\mu = \mu_r < 0$,  while the minimal 
eigenvalue,  say $\gamma_1(\mu)$,  may have an infinite discontinuity
jumping from negative infinity,  at $\mu \to \mu_r - 0$,  to positive
infinity,  at $\mu \to \mu_r + 0$. \\

To show the property (i), we consider the asymptotic limit of ${\bf A}(\mu)$ as
$\mu \to - \infty$. In this limit, the eigenvalues $\gamma(\mu)$ can be
expressed from the algebra of quadratic forms,
\begin{equation}
\label{asymptotic}
\gamma(\mu) = \frac{1}{\mu \; \langle \mbox{\boldmath $\nu$} |
\mbox{\boldmath $\nu$} \rangle}
\left( \sum_{\mu_r < 0} b_r + \sum_{\mu_r > 0} b_r \right),
\end{equation}
where
\begin{equation}
\label{coeffic}
b_r = \left| \sum_{n=1}^N \nu_n \langle \mbox{\boldmath $\psi$}_r | 
\Phi_n {\bf e}_n \rangle \right|^2 \geq 0.
\end{equation}
Since all $b_r$ may not vanish simultaneously for $\mbox{\boldmath $\nu$} 
\neq {\bf 0}$, the eigenvalues $\gamma(\mu)$ are negative definite in Eq. (\ref{asymptotic}) so that $\gamma(\mu) \to -0$ as $\mu \to -\infty$.

To show the property (ii), we take the derivative of the system 
${\bf A} \mbox{\boldmath $\nu$} = \gamma(\mu) \mbox{\boldmath $\nu$}$ 
and use the algebra of quadratic forms. The derivative of $\gamma(\mu)$ 
is then defined for $\mu \leq 0$ excluding the resonant planes at 
$\mu = \{ \mu_{-n(\Lambda)}, \mu_{-n(\Lambda) + 1}, ..., \mu_{-1} \}$ as 
\begin{equation}
\label{derivative}
\frac{d \gamma(\mu)}{d \mu} = \frac{1}{\langle \mbox{\boldmath $\nu$}
| \mbox{\boldmath $\nu$} \rangle}
\langle \mbox{\boldmath $\nu$} | \frac{d {\bf A}(\mu)}{d \mu}
\mbox{\boldmath $\nu$} \rangle =
- \frac{1}{\langle \mbox{\boldmath $\nu$} | \mbox{\boldmath $\nu$} \rangle}
\left( \sum_{\mu_r < 0} \frac{b_r}{(\mu - \mu_r)^2} +
\sum_{\mu_r > 0} \frac{b_r}{(\mu - \mu_r)^2} \right),
\end{equation}
where $b_r$ are defined by the same relation (\ref{coeffic}). Since the 
derivative of $\gamma(\mu)$ is hegative definite in Eq. (\ref{derivative}), 
all eigenvalues $\gamma(\mu)$ are decreasing functions of $\mu$
whenever $d \gamma(\mu)/d \mu$ exists. 

To show the property (iii), we consider the behavior of the eigenvalues 
$\gamma(\mu)$ at the resonant plane $\mu = \mu_r < 0$. In this limit, 
the matrix elements $A_{nm}(\mu)$ have the following asymptotic form,
$$
A_{nm}(\mu) \to \frac{ \langle \Phi_n {\bf e}_n |
\mbox{\boldmath $\psi$}_r \rangle \; \langle 
\mbox{\boldmath $\psi$}_r | \Phi_m {\bf e}_m \rangle}{(\mu - \mu_r)}.
$$
Therefore, the matrix ${\bf A}(\mu)$ has ($N-1$) zero eigenvalues $\gamma(\mu)$ and a single  non-zero eigenvalue $\gamma_1(\mu)$ with the asymptotic value,
\begin{equation}
\label{mineigen}
\gamma_1(\mu) \to \frac{1}{(\mu - \mu_r)} \sum_{n=1}^N
\left| \langle \Phi_n {\bf e}_n | \mbox{\boldmath $\psi$}_r \rangle \right|^2.
\end{equation}
If the sum in Eq. (\ref{mineigen}) does not vanish, the eigenvalue
$\gamma_1(\mu)$ has an infinite discontinuity described in (iii) and, 
according to the property (ii), it is the minimal eigenvalue.  
Other ($N-1$) eigenvalues are in fact non-zero in the limit $\mu \to \mu_r$. 
Since the matrix ${\bf A}(\mu)$ is a meromorphic function of $\mu$ as 
$\mu \leq 0$, the eigenvalue $(\mu - \mu_r) \gamma(\mu)$
is of order of $O(\mu - \mu_r)$ for ($N-1$) non-singular eigenvalues. 
Therefore, the values of $\gamma(\mu)$ are generally non-zero
in the limit $\mu \to \mu_r$. \\

Thus, we have a clear picture how the eigenvalues $\gamma(\mu)$ behave
as functions of $\mu$ [see Figs 1(a,b)].  Starting with small negative 
values as $\mu \to - \infty$, all eigenvalues decrease as $\mu$ grows 
towards the $n(\Lambda)$ resonant planes. At each of those planes, ($N-1$) 
eigenvalues remain continuously decreasing, while one (minimal) 
eigenvalue jumps to a positive domain unless the condition
\begin{equation}
\label{rescond}
\sum_{n=1}^N \left| \langle \Phi_n {\bf e}_n |
\mbox{\boldmath $\psi$}_r \rangle \right|^2 = 0
\end{equation}
is satisfied (this condition will be discussed below). Assuming that 
the condition (\ref{rescond}) is not met, we come to the conclusion
that a root of $\gamma(\mu)$ may occur only after a jump of $\gamma(\mu)$ 
at a resonant plane $\mu = \mu_r$ to a large positive
value,  and further decrease of $\gamma(\mu)$ as $\mu \; (>\mu_r)$ 
grows. The root of $\gamma(\mu)$, if exists for $\mu \leq 0$, 
produces a legitimate solution ${\bf u}_k({\bf x})$ of the problem 
(\ref{auxillary}) under the constraints (\ref{constr}). 
This solution ($\mu, {\bf u}_k$) would then be associated 
with an unstable eigenvalue $\lambda$,  according to the connection
formula (\ref{connection}). Thus, our main task is to control the
behavior of positive $\gamma(\mu)$ in between the plane $\mu = 0$ and
the resonant planes $\mu = \{ \mu_{-n(\Lambda)}, \mu_{-n(\Lambda) + 1},
..., \mu_{-1} \}$.

At the plane $\mu = 0$, the problem (\ref{auxillary}) has a simple
solution for ${\bf u}_k({\bf x})$,
\begin{equation}
\label{stationary}
{\bf u}_{\mu = 0}({\bf x}) = \sum_{n=1}^N \nu_n \frac{\partial
{\bf \Phi}({\bf x})}{\partial \beta_n}.
\end{equation}
Substituting Eq. (\ref{stationary}) into the constraints (\ref{constr}), we
find that ${\bf A}(0) = {\bf U}$, where the matrix ${\bf U}$ is the
Hessian of the energetic surface $\Lambda_s({\bf \beta})$ with the
elements $U_{nm}$ defined by Eq. (\ref{Hessian}). We can now use 
this construction and prove the main results (i)--(iii). In the analysis below 
we assume that the condition (\ref{rescond}) is never met and the 
root of $\gamma(\mu)$ at $\mu \leq 0$ is associated with the 
unstable eigenvalue $\lambda$ of the stability problem (\ref{eigenvalue}). 

The roots of $\gamma(\mu)$ may appear only to the right of any of the
$n(\Lambda)$ resonant plane. There are totally $n(\Lambda)$ jumps of 
$\gamma(\mu)$ to positive values at $\mu \leq 0$ and, therefore,
 no more than $n(\Lambda)$ roots of $\gamma(\mu)$ may exist
for $\mu \leq 0$. 

If $n(\Lambda) = N$, the positive eigenvalues $\gamma(\mu)$
remain continuous after passing the corresponding resonant plane at
$\mu = \mu_r \; (< 0)$. Therefore, the sign of these eigenvalues is
controlled by the eigenvalues of the Hessian matrix ${\bf U}$ at
$\mu = 0$. If $p(U) = N = n(\Lambda)$, all positive eigenvalues
$\gamma(\mu)$ remain positive for $\mu_r < \mu \leq 0$ and no
roots of $\gamma(\mu)$ exist for $\mu \leq 0$ [see Fig. 1(a)]. 
If $p(U) < N = n(\Lambda)$, there exist  $N-p(U)$ negative or 
zero-becoming-negative eigenvalues of ${\bf U}$ that correspond 
to  $N-p(U)$ roots of $\gamma(\mu)$ for $\mu \leq 0$.

If $n(\Lambda) < N$,  then $N - n(\Lambda)$  eigenvalues $\gamma(\mu)$
do not have jumps at the corresponding resonant planes $\mu = \mu_r$. 
They continue to be negative and match at $\mu = 0$ with the
$N-n(\Lambda)$  negative eigenvalues of ${\bf U}$. From this, 
we come to the conclusion that $p(U)$ and $n(U)$ satisfy the constraints 
(\ref{restrict}), i.e. $n(U) \geq N - n(\Lambda)$
or, equivalently, $p(U) \leq n(\Lambda)$. Furthermore, 
other $n(\Lambda)$ $(<N)$ eigenvalues $\gamma(\mu)$ may have 
roots for $\mu \leq 0$ which are completely controlled by the 
remaining $n(\Lambda)$ eigenvalues of ${\bf U}$ according to 
the same criterion as in the case $p(\Lambda) = N$. For instance, 
if $p(U) < n(\Lambda)$, then $n(\Lambda) - p(U)$ negative or 
zero-becoming-negative eigenvalues of 
the matrix ${\bf U}$ correspond to $n(\Lambda) - p(U)$ roots of $\gamma(\mu)$ 
at $\mu \leq 0$. 

If $n(\Lambda) > N$, then $n(\Lambda) - N$ eigenvalues
$\gamma(\mu)$ jump twice in the domain $\mu \leq 0$ leading 
to at least $n(\Lambda) - N$ unconditional roots for $\mu \leq 0$ 
[see Fig.1(b)]. After the jumps, the $N$ eigenvalues $\gamma(\mu)$ 
match the $N$ eigenvalues of the matrix ${\bf U}$ and may have additional 
roots of $\gamma(\mu)$ if $p(U) < N$. Total number of roots of $\gamma(\mu)$ at 
$\mu \leq 0$ is then defined as $(n(\Lambda) - N) + (N - p(U)) = n(\Lambda) - p(U)$. 

The analysis above is valid for the non-degenerate case when the
condition (\ref{rescond}) is never satisfied. However, the 
stability and instability results
(i)-(iii) are not affected even if the condition (\ref{rescond}) is
satisfied for a particular resonant plane $\mu = \mu_r \;(< 0)$. 
In this case, the eigenfunction
${\bf u}_k({\bf x})$ of the operator ${\bf L}_1$ satisfies all
the constraints (\ref{constr}) identically and, therefore, the
eigenvalue $\mu = \mu_r$ is associated with an unstable eigenvalue $\lambda$, 
according to Eq. (\ref{connection}). Although the eigenvalue
$\gamma_1(\mu)$ has no jump at $\mu = \mu_r$ [see Eq. (\ref{mineigen})]
and is continuous, it is still controlled by the negative eigenvalues
of ${\bf U}$ at $\mu = 0$. Indeed, in this case, the minimal eigenvalue
$\gamma_1(\mu)$ at $\mu < \mu_r$ remains negative for $\mu > \mu_r$ and 
matches with a negative
eigenvalue of the matrix ${\bf U}$ (if no other jumps occur in the 
domain $\mu \leq 0$). This additional negative eigenvalue $\mu$ 
still predicts the instability, according to the result (iii). 

Finally, we prove the result (iv) for the instability bifurcation of
multi-component solitary waves. Provided the number $n(\Lambda)$ 
is fixed, the instability bifurcation may only occur when
${\bf A}(0) = {\bf U}$ has a zero eigenvalue for a certain
eigenvector $\mbox{\boldmath $\nu$} = \mbox{\boldmath $\nu$}^{(k)}$.
Let us define ${\bf U} = {\bf U}_{\rm thr}$ at the marginal stability curve
so that the determinant of ${\bf U}_{thr}$ vanishes. The 
instability bifurcations of multi-component solitons were 
considered in Refs. \cite{P3,skryabin} but the results do not agree with each other.  Here, we recover the results of Ref. \cite{P3} and derive the normal 
form (\ref{particle}) by an elegant reduction of general 
algebraic expressions.

Assuming $\mu = 0$ for $\mbox{\boldmath $\nu$} = \mbox{\boldmath $\nu$}^{(k)}$
so that ${\bf U}_{thr} \mbox{\boldmath $\nu$}^{(k)} = {\bf 0}$, we find
the asymptotic solution of Eq. (\ref{eigenvalue}) in the form
(\ref{stationary}) and
\begin{equation}
\label{kinetic}
{\bf w}_{\mu = 0}({\bf x}) = \lambda \sum_{n=1}^N \nu_n^{(k)} {\bf L}_0^{-1}
\frac{\partial {\bf \Phi} ({\bf x})}{\partial \beta_n}.
\end{equation}
In this limit, the second variation $\delta^2 \Lambda$ of the Lyapunov
functional can be found from Eqs. (\ref{second}), (\ref{constr}),
(\ref{stationary}), and (\ref{kinetic}) as follows
\begin{equation}
\label{system}
\delta^2 \Lambda = D_1 \lambda^2,
\end{equation}
where
\begin{equation}
\label{D1}
D_1 = \int_{-\infty}^{\infty} d {\bf x} \sum_{m=1}^N \frac{1}{\Phi_m^2({\bf
x})}
\left[ \sum_{n=1}^N \nu_n^{(k)} \int_0^{\bf x} d {\bf x}' \Phi_m({\bf x}')
\frac{\partial \Phi_m({\bf x}')}{\partial \beta_n} \right]^2.
\end{equation}
The integral converges under the condition that $\mbox{\boldmath $\nu$}^{(k)}$
is a solution of the equation ${\bf U}_{\rm thr} \mbox{\boldmath $\nu$}^{(k)} 
= {\bf 0}$. On the other hand, the perturbation (\ref{stationary}) shifts the soliton parameter $\mbox{\boldmath $\beta$}$ according to the expression,
${\bf \Phi}({\bf x},\mbox{\boldmath $\beta$}) + {\bf u}_{\mu = 0}({\bf x})
\to {\bf \Phi}({\bf x},\mbox{\boldmath $\beta$} + \mbox{\boldmath
$\nu$}^{(k)})$. As a result, the second variation can be closed as
\begin{equation}
\label{close}
\delta^2 \Lambda = 2 \left[ \Lambda - \Lambda_s(\mbox{\boldmath $\beta$}
+ \mbox{\boldmath $\nu$}^{(k)}) \right] \to - D_0,
\end{equation}
where
\begin{equation}
\label{D0}
D_0 = \langle \mbox{\boldmath $\nu$}^{(k)} | {\bf U}
\mbox{\boldmath $\nu$}^{(k)} \rangle.
\end{equation}
The parameter $\Lambda$ in Eq. (\ref{close}) is chosen from the condition
that the first variation of $\Lambda_s(\mbox{\boldmath $\beta$} +
\mbox{\boldmath $\nu$}^{(k)})$ vanishes for arbitrary $\mbox{\boldmath
$\nu$}^{(k)}$. This gives the connection formula: $\Lambda \equiv
\Lambda_{st} = H_s(\mbox{\boldmath $\beta$}) + \sum_{n=1}^N
(\beta_n + \nu_n^{(k)}) Q_{sn}(\mbox{\boldmath $\beta$})$. Equating
Eq. (\ref{system}) and Eq. (\ref{close}), we recover the result of the bifurcation theory,
\begin{equation}
\label{bifurcation}
\lambda^2 = - \frac{D_0}{D_1}.
\end{equation}
Since $D_1 > 0$ [see Eq. (\ref{D1})], the positive values of $\lambda^2$
occur when the determinant of the matrix ${\bf U}$ is small and negative
(i.e. the matrix ${\bf U}$ has a zero-becoming-negative eigenvalue when 
the soliton parameter $\mbox{\boldmath $\beta$}$ crosses the marginal stability 
curve). The explicit formulas of the soliton bifurcation theory
provide an alternative and more compact form for the
determinants $D_0$ and $D_1$ compared to those obtained in Ref. \cite{skryabin}.

The normal form (\ref{particle}) follows from Eqs. (\ref{system}) and
(\ref{close}) when $\Lambda = \Lambda_{st} + E$, and the perturbation
vector $\mbox{\boldmath $\nu$}^{(k)}$ is replaced by a slowly varying
vector $\mbox{\boldmath $\nu$} = \mbox{\boldmath $\nu$}(z)$ (see \cite{P3} 
for details). Then, the
surface $\Lambda_s(\mbox{\boldmath $\beta$} + \mbox{\boldmath $\nu$})$
is extended beyond the second variation limit, and the linear approximation
is converted into the slope: $\lambda \mbox{\boldmath $\nu$}(z) =
d \mbox{\boldmath $\nu$}(z)/d z$. The mass constants $M_{nm}$ follow from Eq. (\ref{D1}) in the explicit form:
\begin{equation}
\label{mass}
M_{nm} = \int_{-\infty}^{\infty} d{\bf x} \sum_{k=1}^N
\frac{1}{\Phi_k^2({\bf x})} \left( \int_0^{\bf x} d {\bf x}'
\Phi_k({\bf x}') \frac{\partial \Phi_k({\bf x}')}{\partial \beta_n} \right)
\left( \int_0^{\bf x} d {\bf x}' \Phi_k({\bf x}') \frac{\partial
\Phi_k({\bf x}')}{\partial \beta_m} \right).
\end{equation}
The normal form (\ref{particle}) resembles the conserved sum of the
kinetic energy and potential energy $W(\mbox{\boldmath $\beta$},\mbox{\boldmath $\nu$})$ of a particle moving in a $N$-dimensional space.  
We notice that the kinetic energy with the ``mass'' matrix (\ref{mass})
is positive definite and the unperturbed multicomponent solitary wave 
(i.e. that with $\mbox{\boldmath $\nu$} = {\bf 0}$) is a stationary point 
of $W(\mbox{\boldmath $\beta$},\mbox{\boldmath $\nu$})$ for any 
$\mbox{\boldmath $\nu$}$. Thus, the stability of multi-component solitons
resembles the stability of a particle located at an equilibrium point of
the $N$-dimensional field \cite{P3}. Under the condition that $n(\Lambda) = N$, the particle is {\em stable} if in the $\mbox{\boldmath $\beta$}$-space the potential energy surface $W(\mbox{\boldmath $\beta$})$ is concave up,  and it is {\em unstable} if the potential energy surface is saddle-type or concave down. If $n(\Lambda) < N$, the potential energy surface $W(\mbox{\boldmath $\beta$})$
always has some $N - n(\Lambda)$ negative directions that do not affect
the stability properties of the particle. However, the remaining $n(\Lambda)$
$(< N)$ directions of the potential energy surface define the stability
of the particle with the same criterion as above. Finally, for the case
$n(\Lambda) > N$, the soliton stability properties defined by the type of the potential energy surface $W(\mbox{\boldmath $\beta$})$ are not conclusive since 
the corresponding unstable eigenvalues coexist with additional
$n(\Lambda) - N$ unconditionally unstable eigenvalues.

\section{Example: two coupled NLS equations}

In order to demonstrate how our general theory can be applied 
to a particular physical problem and also to compare the 
stability and instability results (ii)-(iii) with some earlier 
known examples, we consider here an important case of 
{\em two incoherently coupled  NLS equations} in (1+1) 
dimension (see, e.g., \cite{berge,Yew,Yang}):
\begin{eqnarray}
\nonumber
i \frac{\partial \psi_1}{\partial z} +
\frac{\partial^2 \psi_1}{\partial x^2} +
\left( |\psi_1|^2 + \gamma |\psi_2|^2 \right) \psi_1 & = & 0,  \\
\label{couple}
i \frac{\partial \psi_2}{\partial z} +
\frac{\partial^2 \psi_2}{\partial x^2} +
\left( |\psi_2|^2 + \gamma |\psi_1|^2 \right) \psi_2 & = & 0,
\end{eqnarray}
where $\gamma$ is a coupling parameter. System (\ref{couple}) 
is a two-component reduction of the general $N-$component 
system (\ref{NLS}) for $d_1 = d_2 = 1$, $\gamma_{11} = 
\gamma_{22} = 1$, and $\gamma_{12} = \gamma_{21} = \gamma$.
An explicit soliton solution can be easily found  
for $\beta_1 = \beta_2 = \beta$ and $\gamma > -1$ in the form,
\begin{equation}
\label{equalamplit}
\Phi_1(x) = \Phi_2(x) = \sqrt{\frac{2 \beta}{1 + \gamma}} \; {\rm sech}
\left( \sqrt{\beta} x \right).
\end{equation}
This solution describes a two-component solitary wave with the components of equal  amplitudes.  It corresponds to a straight line $\beta_1 = \beta_2$ in the 
parameter plane ($\beta_1,\beta_2$) of a general {\em two-parameter 
family of solitary waves} of the model (\ref{couple}).  
When $-1 < \gamma \leq 0$, such two-parameter solitons may exist 
everywhere in the plane ($\beta_1,\beta_2$), while for $\gamma > 0$, 
the soliton existence domain  is restricted by two bifurcation curves, 
$\beta_2 = \omega_{\pm}(\gamma) \beta_1$, where
\begin{equation}
\label{domain}
\omega_{\pm}(\gamma) = \left( 
\frac{\sqrt{1 + 8 \gamma} - 1}{2} \right)^{\pm 2}.
\end{equation}

Approximate analytical expressions can also be obtained in the vicinity of 
the bifurcation curves (\ref{domain}), when one of the components of 
a composite solitary wave becomes small, while the other one is 
described by the scalar NLS equation. Such a case, when one of the 
component creates an effective waveguide  that guides the other 
component, is known to describe the so-called {\em shepherding effect} 
where the large-amplitude  component plays a role of a shepherding 
pulse \cite{bpw}. The composite soliton,  that describes a shepherding 
pulse $\psi_1$ guiding a small component pulse $\psi_2$,  can be 
found in the form (see also Ref. \cite{Yang}), 
\begin{equation}
\label{decoupled}
\Phi_1 = R_0(x) + \epsilon^2 R_2(x) + {\rm O}(\epsilon^4), \;\;\;\;
\Phi_2 = \epsilon S_1(x) + {\rm O}(\epsilon^3). 
\end{equation}
It exists in the vicinity of the bifurcation curve, 
\begin{equation}
\label{bif}
\beta_2 = \omega_+(\gamma) \beta_1 + \epsilon^2 
\omega_{2+}(\gamma) \beta_1 + {\rm O}(\epsilon^4),
\end{equation} 
and the main terms of the asymptotic series (\ref{decoupled}), 
(\ref{bif}) are defined as 
$$
R_0 = \sqrt{2 \beta_1} {\rm sech} \left(\sqrt{\beta_1} x \right), \;\;\;\;
S_1 = \sqrt{\beta_1} {\rm sech}^{\sqrt{\omega_+}} 
\left( \sqrt{\beta_1} x \right),
$$
and 
$$
\omega_{2+} = \frac{ \int_{-\infty}^{\infty} dx \left( S_1^4 + 
2 \gamma R_0 R_2 S_1^2 \right)}{\int_{-\infty}^{\infty} dx S_1^2}, 
$$
The second-order correction $R_2(x)$ is a solution 
of the differential equation,
$$
\left[ - \partial_x^2 + \beta_1 - 6 \beta_1 \; {\rm sech}^2 \left( 
\sqrt{\beta_1} x \right) \right] R_2 = \gamma R_0 S_1^2.
$$

From the domain of existence of the two-component soliton,  it follows 
that $\omega_{2+}(\gamma) > 0$,  for $0 < \gamma < 1$,  and 
$\omega_{2+}(\gamma) < 0$,  for $\gamma > 1$. At $\gamma = 1$ 
(the so-called integrable Manakov case), a family of two-parameter composite
solitons becomes degenerated: it exists on the line $\beta_1 = \beta_2$ 
but,  generally, it is different from the one-parameter solution 
(\ref{equalamplit}).  The coupled solitons are known to be stable for 
the integrable case  $\gamma = 1$. Here we apply the stability theory 
developed above  and prove that the (1+1)-dimensional two-parameter 
solitons, including the solitons of equal amplitudes (\ref{equalamplit}), 
are {\em stable} for $\gamma \geq 0$, and {\em unstable} for $\gamma < 0$. 

First, we evaluate the indices $p(U)$ and $n(\Lambda)$ for the 
explicit solution (\ref{equalamplit}). As follows from Eqs. (\ref{couple}) 
and (\ref{equalamplit}),  the Hessian matrix ${\bf U}$ with 
the elements (\ref{Hessian}) can be found in the form,
$$
\frac{\partial Q_1}{\partial \beta_1} = 
\frac{\partial Q_2}{\partial \beta_2}
= \frac{1}{\sqrt{\beta} ( 1 + \gamma)} \;\;\;\mbox{and}\;\;\;
\frac{\partial Q_1}{\partial \beta_2} = 
\frac{\partial Q_2}{\partial \beta_1}
= -\frac{\gamma}{\sqrt{\beta} ( 1 + \gamma)}.
$$
As follows from these results,  the Hessian matrix has $p(U) = 2$ 
positive eigenvalues,  for $-1 < \gamma < 1$, and $p(U) = 1$ positive 
eigenvalue,  for $\gamma > 1$. On the other hand, the linear matrix 
operator ${\bf L}_1$ given below Eq. (\ref{second}) can be diagonalized for
linear combinations of the eigenfunctions,  $v_1 = u_1 + u_2$ and $v_2 = u_1 - u_2$, such that 
\begin{eqnarray}
\nonumber
\left[ - \partial^2_x + \beta - 6 \beta \; {\rm sech}^2
\left(\sqrt{\beta} x \right) \right] v_1 & = & \mu v_1, \\
\label{diagonal}
\left[ - \partial^2_x + \beta - 2 \beta 
\frac{(3 - \gamma)}{(1 + \gamma)} \; 
{\rm sech}^2 \left(\sqrt{\beta} x \right) \right] v_2 & = & \mu v_2.
\end{eqnarray}
Both the operators in Eqs. (\ref{diagonal}) are the linear 
Schr\"{o}dinger operators with solvable sech-type 
potentials, and the corresponding eigenvalue spectra are 
well studied. The first operator always has a single 
negative eigenvalue for $\mu = -3 \beta$, whereas the second 
operator  has no negative eigenvalues, for $\gamma > 1$, has 
a single negative eigenvalue, for $0 < \gamma < 1$, and it has 
two negative eigenvalues, for $-1 < \gamma < 0$. Thus, in total there 
exist $n(\Lambda) = 3$ negative eigenvalues,  for 
$-1 < \gamma < 0$, $n(\Lambda) = 2$ negative eigenvalues,  
for $0 < \gamma < 1$, and $n(\Lambda) = 1$ negative eigenvalue,  
for $\gamma > 1$.

Applying the stability and instability results (ii)-(iii) 
obtained and discussed in Secs.  II  and III, we come to the conclusion
 that  the soliton solution (\ref{equalamplit}) with equal amplitudes 
is {\em linearly stable} for $\gamma > 0$, since in this domain 
$p(U) = n(\Lambda) = \{1,2\}$, and it is {\em linearly unstable} 
for $-1 < \gamma < 0$, since in this domain 
$p(U) = 2 < n(\Lambda) = 3$. 

The soliton stability in the model (\ref{couple}) for $\gamma > 0$ 
was also studied  by  Berg\'e \cite{berge} who considered the case
of degenerate one-parametric solitary waves (\ref{equalamplit}). 
Here we have extended those results to a general case: 
the same stability and instability results are valid for 
the two-parameter family of solitons provided that the 
indices $p(U)$ and $n(\Lambda)$ remain unchanged for the values 
($\beta_1,\beta_2$) that do not belong to the line $\beta_1 = \beta_2$. 
Indeed, applying a perturbation theory for small $\gamma$ 
(see \cite{Yang} for details), one can show that $n(\Lambda) = 3$,  
for $-1 \ll \gamma < 0$,  and $n(\Lambda) = 2$,  for $0 < \gamma \ll 1$. 

Finally, we consider the other limiting case that describes 
the shepherding effect [see  Eqs. (\ref{decoupled}) and (\ref{bif})]. 
In this limit, the elements (\ref{Hessian}) of the  Hessian matrix 
${\bf U}$ can also be calculated in explicit analytical form, 
$$
\frac{\partial Q_1}{\partial \beta_1} = \frac{1}{\sqrt{\beta_1}} 
+ \frac{r_2^2}{\omega_{2+} s_1} + {\rm O}(\epsilon^2), \;\;\;\;
\frac{\partial Q_1}{\partial \beta_2} = 
\frac{\partial Q_2}{\partial \beta_1} = 
\frac{r_2}{\omega_{2+}} + {\rm O}(\epsilon^2), \;\;\;\;
\frac{\partial Q_2}{\partial \beta_2} = 
\frac{s_1}{\omega_{2+}},
$$
where 
$$
s_1 = \frac{1}{2} \int_{-\infty}^{\infty} S_1^2 dx \;\;\;
\mbox{    and    } \;\;\;
r_2 =  \int_{-\infty}^{\infty} R_0 R_2 dx. 
$$
Since $s_1 > 0$ for any $\gamma$, while $\omega_{2+}(\gamma) > 0$ 
for $0 < \gamma < 1$ and $\omega_{2+}(\gamma) < 0$ for $\gamma > 1$, 
the Hessian matrix ${\bf U}$ calculated for the shepherding soliton 
(\ref{decoupled}) has $p(U) = 2$ positive eigenvalues, 
for $0 < \gamma < 1$, and $p(U) = 1$ positive eigenvalue, 
for $\gamma > 1$. 

On the other hand, the linear matrix operator ${\bf L}_1$ 
cannot be diagonalized for the shepherding 
soliton (\ref{decoupled}) unless $\epsilon = 0$. In the latter 
(decoupled) case, it has a single negative eigenvalue at 
$\mu = - 3 \beta_1$ and a double degenerate zero eigenvalue. 
When $\epsilon \neq 0$, the zero eigenvalue shifts to become
 $\mu = - 2 \omega_{2+}(\gamma) \beta_1 \epsilon^2 + 
{\rm O}(\epsilon^4)$. 
Therefore, the matrix operator ${\bf L}_1$ for the shepherding 
soliton (\ref{decoupled}) has $n(\Lambda) = 2$ negative 
eigenvalues,  for $0 < \gamma < 1$,  and $n(\Lambda) = 1$ negative 
eigenvalue,  for $\gamma > 1$. Thus, we come to the conclusion that 
{\em the  shepherding soliton is stable} for $\gamma > 0$ since 
$p(U) = n(\Lambda) = \{1,2\}$. 

\section{Conclusion}

We have developed a rigorous stability analysis of multi-component 
solitary waves considering a system of incoherently coupled NLS 
equations (\ref{NLS}) as a particular but important physical example.  
The method and, correspondingly, both stability and instability 
results can be extended to other types of solitary waves, such as 
multi-component spatial solitons  (e.g., incoherent solitons) in 
non-Kerr (e.g., saturable) media, parametric solitary waves in 
quadratic (or $\chi^{(2)}$) optical  media, etc. In all such cases, 
our stability and instability results (i)--(iv) can be readily 
generalized with {\em a rigorous proof} of some of the previously 
known results of the asymptotic multi-scale expansion theory. 
However, additional analysis is required in each of those cases 
in order to clarify the conditions when these results 
completely define the stability properties of multi-component 
solitary waves. In the cases beyond these conditions, oscillatory 
instabilities may occur, and appropriate studies should rely solely 
on the  numerical analysis of the corresponding eigenvalue 
problems and their linear spectra.

\section*{Acknowledgments} 

The authors appreciate collaboration with A.A. Sukhorukov and 
E.A. Ostrovskaya.  Dmitry Pelinovsky thanks C.K.R.T. Jones, 
K. Promislow, M. Weinstein, J. Yang, and A. Yew for stimulating 
discussions at different stages of the work. The work is supported 
by the Performance and Planning Fund of the Institute of Advanced 
Studies and by the Australian Photonics Cooperative Research Centre.

\begin{center}
\bf Figure Captions
\end{center}

{\bf Fig. 1:} Eigenvalues $\gamma$ versus $\mu$ in the problem 
${\bf A}(\mu) \mbox{\boldmath $\nu$} = \gamma(\mu) 
\mbox{\boldmath $\nu$}$ for $N=3$: (a) a stable problem with no 
roots of $\gamma(\mu)$ for $\mu \leq 0$, when $p(U) = n(\Lambda) = 3$; 
(b) an unstable problem with a single root of $\gamma(\mu)$ for 
$\mu \leq 0$, when $p(U) = 3 < n(\Lambda) = 4$. 

\begin{figure}
\begin{center}
\psfig{file=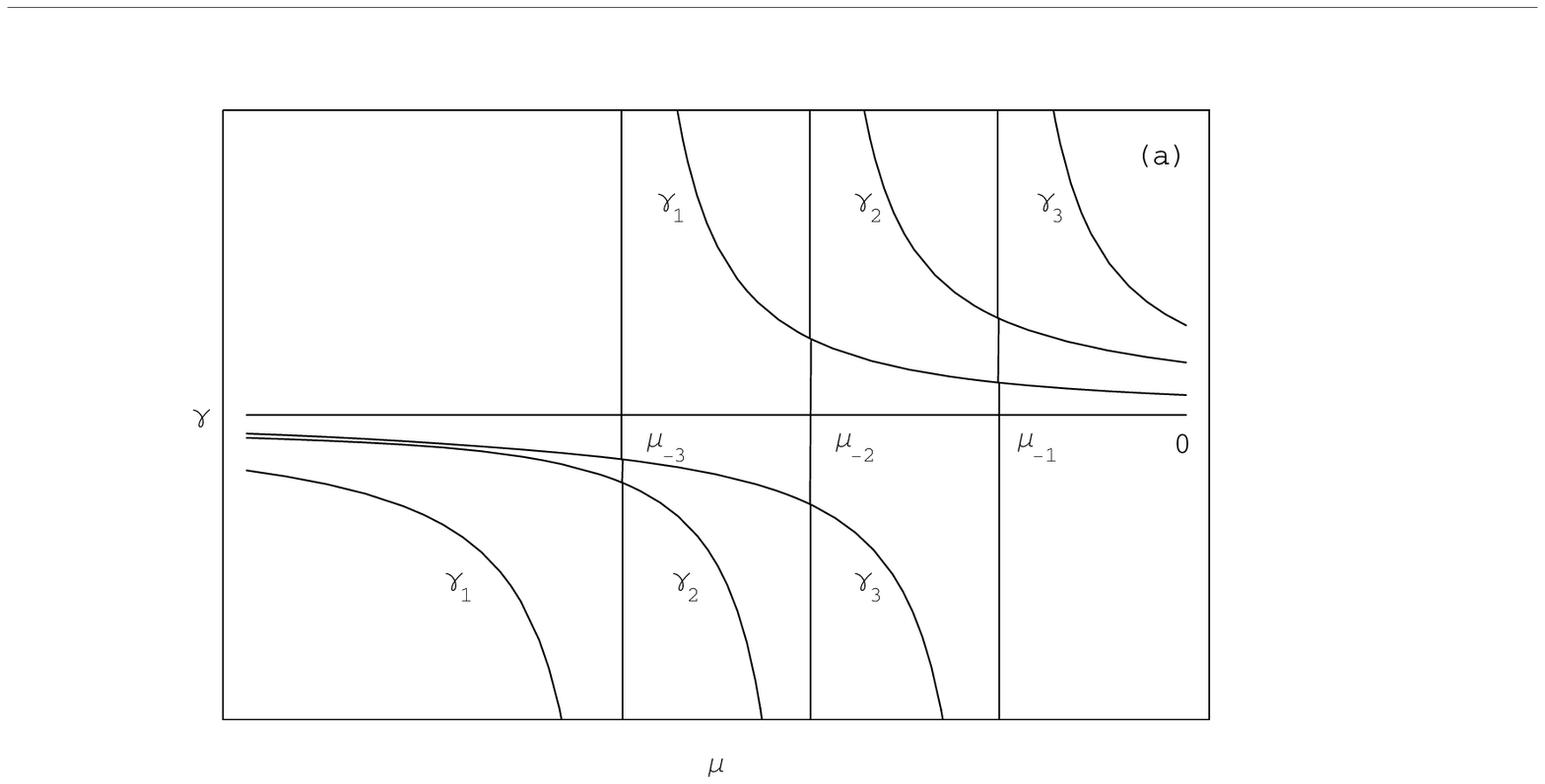, height=25cm, width=20cm}
\end{center}
\end{figure}

\begin{figure}
\begin{center}
\psfig{file=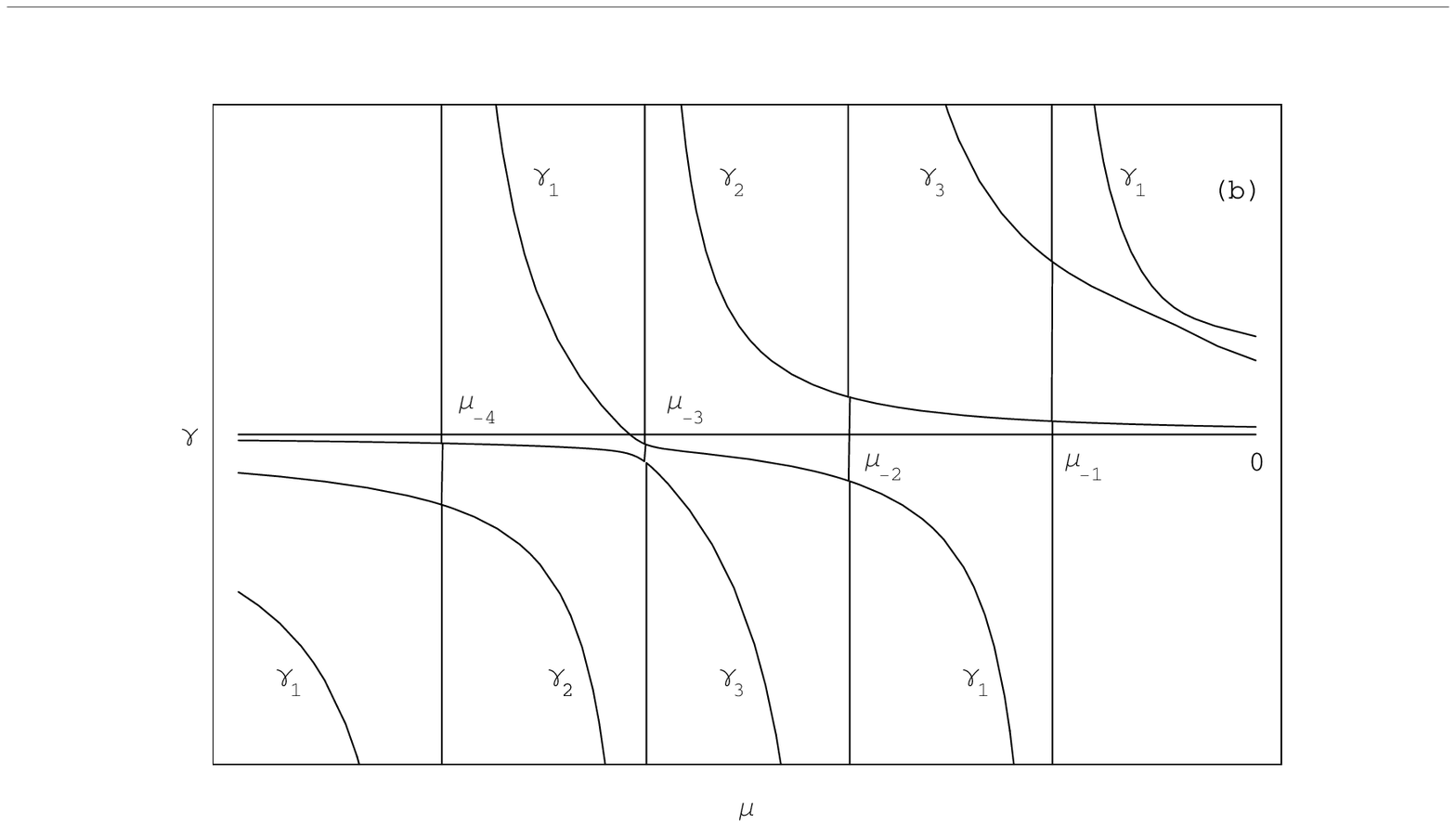, height=25cm, width=20cm}
\end{center}
\end{figure}

\end{document}